# C-slow Technique vs Multiprocessor in designing Low Area Customized Instruction set Processor for Embedded Applications

Muhammad Adeel Akram
COMSATS Insititute of IT
Quaid Avenue, Wah Cantt.
Pujnab, Pakistan

Aamir Khan
COMSATS Insititute of IT
Quaid Avenue, Wah Cantt.
Pujnab, Pakistan

Muhammad Masood Sarfaraz
COMSATS Institute of IT
Quaid Avenue, Wah Cantt
Punjab, Pakistan

**ABSTRACT**

The demand for high performance embedded processors, for consumer electronics, is rapidly increasing for the past few years. Many of these embedded processors depend upon custom built Instruction Ser Architecture (ISA) such as game processor (GPU), multimedia processors, DSP processors etc. Primary requirement for consumer electronic industry is low cost with high performance and low power consumption. A lot of research has been evolved to enhance the performance of embedded processors through parallel computing. But some of them focus superscalar processors i.e. single processors with more resources like Instruction Level Parallelism (ILP) which includes Very Long Instruction Word (VLIW) architecture, custom instruction set extensible processor architecture and others require more number of processing units on a single chip like Thread Level Parallelism (TLP) that includes Simultaneous Multithreading (SMT), Chip Multithreading (CMT) and Chip Multiprocessing (CMP). In this paper, we present a new technique, named C-slow, to enhance performance for embedded processors for consumer electronics by exploiting multithreading technique in single core processors. Without resulting into the complexity of micro controlling with Real Time Operating system (RTOS), C-slowed processor can execute multiple threads in parallel using single datapath of Instruction Set processing element. This technique takes low area & approach complexity of general purpose processor running RTOS.

**Keywords**: Instruction Set Architecture (ISA), Instruction Level Parallelism (ILP), Very Long Instruction Word (VLIW), Thread Level Parallelism (TLP), Simultaneous Multithreading (SMT), Chip Multithreading (CMT), Chip Multiprocessing (CMP)

## 1. INTRODUCTION

From the past few years, consumer electronic industry, toys to high end game consoles and from mp3 players to PDAs and Laptops, is growing with sky rocket speed. One of the primary requirements for these embedded systems is high performance with minimal silicon area cost. Most of consumer based embedded processors require custom built ISAs in contrast to general purpose processors. The ISA of a general purpose processor is designed to meet the requirements in a variety of domains. So a general purpose processor is not suitable to meet the high performance requirement of application specific circuits. Thus Application Specific Instruction Set Processors (ASIPs) are the promising way to enhance the computation performance of consumer based embedded systems [7], [13]. Moreover cost and time to market is also much important for consumer electronics [6] and ASIPs allow high performance with automatic design even for high level architecture. It also `affects the cost as embedded applications are developed in huge quantity, a small decrease in amount effects a lot. ARM [1], a 32 bit reduced Instruction set architecture is a common example.

During the last decade processor clock frequency was synchronous with performance i.e. faster processor means more computing power. During this, the target of most of the performance related research was single core processors. Within a few years, clock speed reaches up to a point where heat dissipation across the chip is at a dangerous level [2]. Thus performance with single processor reaches to its optimal limit. Then processor designers target to design multicore processors to enhance performance as theoretically adding additional core in a chip means double the performance. Some examples of multicore processors are Cell [3], MP211 [4], FR1000 [5]. Most techniques developed, during recent years, target multicore processors.

Most of the performance enhancement techniques exploit parallelism within processors. Two types of parallel computation techniques are common within processors, Instruction Level Parallelism (ILP) [10], [11] and Thread Level Parallelism (TLP) [12]. The basic idea behind both is same because both identify independent instructions and utilize parallel to compute in parallel. Neither of the technique is promising to adopt dynamically hardware changes.

*Instruction Level Parallelism* exploits parallel execution by utilizing independent instructions within a program such as memory load/store, addition/multiplication instructions etc. Normally ILP architectures are transparent to users. One primary difference between ILP processor and normal RISC based machine is that ILP processors require more hardware resources to perform parallel execution. There are two most common processor architectures that exploit instruction level parallelism, VLIW [25] processors and Superscalar processors [26]. Superscalar processors are sequential architectures that in which program do not provide any explicit information about parallelism. So program does not know the presence of dependencies present in a program rather it is the task of hardware to find the dependencies between instructions. The primary issue in superscalar processors is how to execute instructions in parallel. As the superscalar processors support sequential execution, rather to execute multiple instructions in single clock cycle, it is more convenient to increase the clock speed up to number of instructions to execute in parallel and issue only one instruction in a single clock cycle. In superscalar processors, it is known as super pipelining [27]. One major problem in superscalar processors is unpredictable branches that reduce the level of performance. This problem can be solved to some extent with speculative execution in which conditional branches are executed first before their control dependencies branches are issued. Many architectures are designed to support speculative execution [13], [14] but hardware can only support a small amount of parallelism up to fetched instructions. The drawback of superscalar processors is the increase in architecture complexity. Because dependence check, branch predictor, reorder buffer introduce makes architecture complex.

VLIW architecture also exploits instruction level parallelism. In order to exploit parallelism system must have knowledge about the independent instructions present in the program. In spite of superscalar processors, in VLIW processors the compiler identifies the presence of parallelism within a program and forward the information to the hardware about the instructions that are independent to each other. Now the hardware does not need to investigate about dependencies. Hence the scheduling and dependence check is moved from hardware level to compiler level so there is a great affect on circuit complexity. VLIW fetches only one instruction per clock cycle that consists of different operations for different processing units to execute in parallel. ILP supports out of order execution in which instructions are rearranged to reduce the interference between computation and memory reference instructions. Two types of instruction scheduling exist in ILP processors, Static ILP scheduling architectures and Dynamic ILP scheduling architectures. In static architecture scheduling is not performed at run time because processor assume that compiler already schedule the instructions and instructions are issued in the same way as they come to machine program [19],[20]. In dynamic scheduling is performed at run time using dedicated hardware. But the performance of ILP processors is limited to the number of independent instructions present in a program. To enhance the performance of ILP design, there must be more independent instructions in a program that can be executed in parallel.

*Simultaneous Multithreading (SMT)* processors, in contrast to ILP allow multiple independent threads to issue multiple processing units. The main purpose of SMT architecture is the maximum processor utilization in the presence of long memory latencies and limited parallelism present in a single thread. SMT exploits some features of superscalar processors such as to issue multiple instructions and also have hardware support to fetch instructions from multiple threads. Thus SMT is a technique which supports the multiple instructions from multiple threads at the same time, hence exploits both ILP and TLP. In SMT processors parallel threads come from multithreaded, parallel program or from independent multi programs. And instruction level parallelism obtains from a single thread or from a single program. In [9], SMT processor's working is discussed and how they are different from other multithreaded architectures and superscalar processors. There are two types of wastes in any type of processor design, horizontal waste and vertical waste. Horizontal waste occurs when all issue slots in single cycle are not filled with instructions. The amount of Horizontal waste is the number of empty slots. The Vertical waste occurs when there is no instruction issued to any issuing slot i.e. all slots is empty as shown in the figure 1.

Normal superscalar processors exploit instruction level parallelism. They execute multiple parallel instructions from a single thread i.e. instructions that are independent to each other. Thus the performance of superscalar processors certainly depends upon the number of independent instructions present in a thread. On the other hand multithreaded processors fetch all instructions from same thread in a single cycle. In next clock cycle they switch the context to new thread and now execute instructions from this new thread. The primary effectiveness of multithreaded processors is that these can tolerate memory latencies thus reducing the vertical waste but they cannot reduce the horizontal waste because they still depend on the independent instructions present in a thread to reduce horizontal waste.

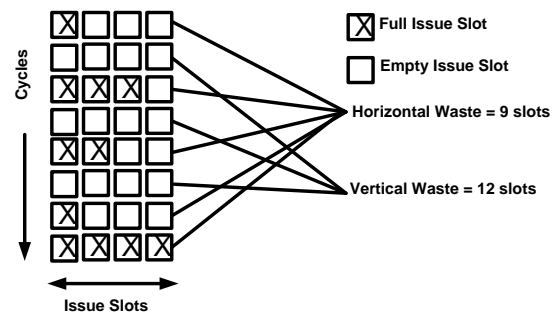

**Fig.1. Horizontal and Vertical waste representation**

Despite all these, SMT processors [24] select instruction from multiple threads in every clock cycle. It schedules the on chip resources in way to maximize the hardware utilization. If more number of independent instructions is available, exploits instruction level parallelism otherwise to utilize hardware instructions from different cycles are selected as shown in the figure 2.

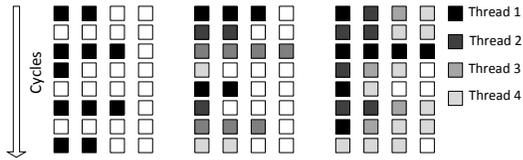

**Fig. 2. (a) Superscalar processor behavior, (b) multithreaded processor and (c) simultaneous multithreaded processor**

In the second section we will discuss some previous work done to design processors especially for embedded applications. In third section we described our proposed model and effects of area and throughput.

## 2. RELATED WORK

There is a lot of work have been evolved to increase the performance of processor. There is large gap between processor's computation speed and memory speed. This is the primary reason in degrading the performance of embedded systems. Many techniques have been evolved to reduce this gap. Pre fetching, multilevel caches, compiler optimization are some of these techniques. To enhance the computation performance of embedded systems, ILP and TLP are two most used approaches. Different models have been proposed to enhance performance by exploiting these two techniques. Some of these models are discussed here.

An architecture is proposed in [16] that exploits both TLP and ILP. To exploit TLP, the design contains much thread processing units and each of these processing units its own program counter, cache memory to perform speculative execution and instruction execution path. It can execute multiple instructions from different threads. But there is a large area because of multiple caches, program counters and ALUs.

Another model proposed in [17], [19], known as EPIC. This model is an enhancement of VLIW but also exploit some features of superscalar processors. Intel's IA-64[18] was first commercially available ISA that was based on EPIC.

SMTA, proposed in [21], consists of a number of thread slots and a thread dispatcher. Each thread slot has its own program counter, instruction and decode unit. The execution results are transferred to another thread if required through communication unit.

Polymorphous Trips Architecture [22] is a single processor core model with a memory system. It supports ILP. Normally it executes instruction in a serial manner but when parallelism is available, it divides itself logically into multiple processing units. But still to exploit parallelism it required more resources.

A stream processor [23] is another approach to enhance performance of embedded systems. Stream processors consist of clusters of functional units. They exploit ILP within a cluster and Data parallelism (DP) [15] between clusters. Each cluster consists of a number of ALUs and use VLIW format by microcontroller. Stream processors also exploit Data Parallelism, a number of data streams that require the same operation to be performed.

## 3. C-SLOW PROCESSOR DESIGN

*Basic concept* of C-slow technique is briefly discussed first. C-slow is a technique that is used to improve the performance especially when there is feedback loops in the design. In the presence of feedback loops, processor cannot exploit the parallelism because next computation is only possible when results of previous computation are available. Hence in these cases we cannot exploit ILP or TLP by simply introducing pipelining registers or using multiple processing units. C-slow was proposed by Leiserson et al. [28] to enhance computation performance in the presence of feedback loops. Figure 3 illustrates the difference between conventional pipelining and C-slow retiming [29].

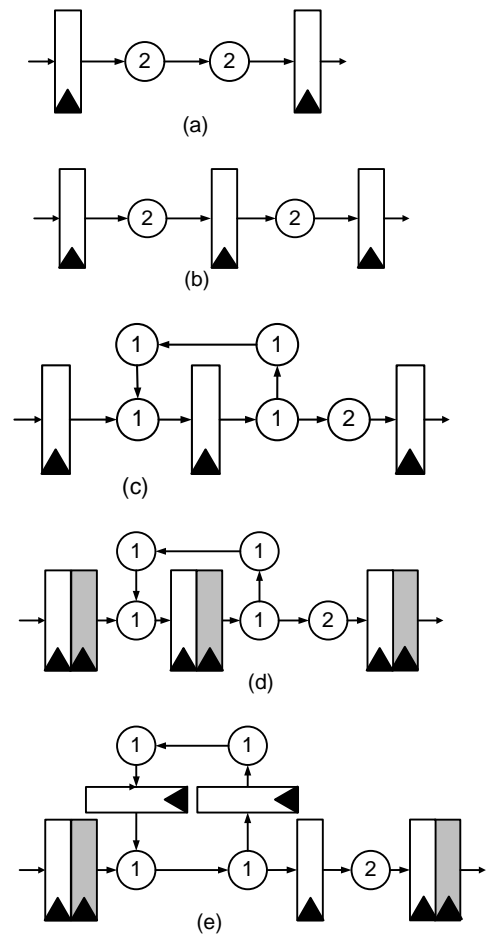

**Fig. 3. (a) A simple feed forward circuit, (b) After pipelining of (a), (c) A circuit with feedback path, (d) After C-slow, (e) After retiming of (d)**

As shown in the figure 3 (a) represents a simple feed forward circuit having critical path 4. After pipelining, critical path is reduced to 2 as shown in (b). A circuit with feedback path is represented in (c). Here feedback is the critical path. After C-slow and retiming circuit is represented in (d) and (e) respectively. Now in every clock cycle, the circuit can process two independent computations by taking data from two independent streams. Hence the input register needs not to wait for the feedback computation results to fetch another input. From above described figure, C-slow is technique that enhances throughput by replacing each register with C number of registers. And C numbers of computations are possible simultaneously.

*C-slow Processor Architecture* consists of registers such ac program counter address register etc. are replaced by C registers as shown in figure 4. All other architecture remains same. The biggest complications in architecture are the implementation of various types of memories. The first type provides the C-slow semantics complete independence, where a thread has a complete independent view. This is applicable to the register file and state registers. In C-slow design, the register file is increased C times.

A hardware thread counter is used to select the group of registers which is being used, so that each can see its own set of registers and all read and writes of different threads are going to different locations.

The second type of memory is completely shared such as main memory and cache memory. Normally these memories are placed out of C-slow portion. There are two ways to access to access cache memory. If the cache is physically addressed, to enhance the throughput just required pipeline the cache so that interlocked read/write instructions has time to be completed. In contrast, when virtually addressed caches are used, they require some way so that one thread cannot access the memory of other thread and require a record to be maintained for virtual to physical address mappings to ensure coherency between threads.

Third one is dynamically shared in which a hardware thread ID or software thread context ID is tagged to each thread. This is best approach for branch predictors and similar caches. These types of memories do not need to be increased in size.

*C-slow as Multithreading* are elaborated. There are numerous architecture proposed to exploit multithreading. All these architectures share same idea: increasing system throughput by executing multiple threads simultaneously. These architectures can be categorized into four classes: always context switching (Heap and Tera [30]), SMT, context switching on event and interleaved multithreading.

The primary idea of C-slow retiming is applicable to highly complex designs such as microprocessors. In those cases, it is not simply a matter of adding registers and balancing delays. The changes in the design are comparatively small than the benefits. The C-slow design produces a simple, statically scheduled, high clock rate, multithreaded design, similar to interleaved multithreaded design. C-slow processor architecture alternates between fixed numbers of threads in a round-robin manner creating the illusion of a multiprocessor system. C-slowing needs three design changes, increasing register file size and modifying TLB, changing cache interface, and modifying interrupt routines.

To create the illusion that each thread is processing on a separate processor i.e. multiple threads are executing on multiple processors, each thread has its own translational memory. This can be performed by increasing size of TLB by C times so that each thread accesses its own set.

## 4. SIMULATION RESULTS

In order to check the performance of proposed C-slow processor over simple processor, a simple instruction set architecture is designed as given in appendix. The given ISA is implemented on *SPARTAN 3* FPGA to test the performance of C-slow processor against simple design. And check the results of execution of multiple threads from different applications. When C number of threads is executed on a simple pipelined design, total number of clock cycles is equal to the sum of clock cycles required to execute C number of single threads. When the same test is performed on *C-Slow*-based design, total number of clock cycles required to execute C number of threads is equal to the thread which requires maximum clock cycles. Fig.5 represents the execution results of one; two and three threads respectively on simple pipelined design and *C-slow* based design.

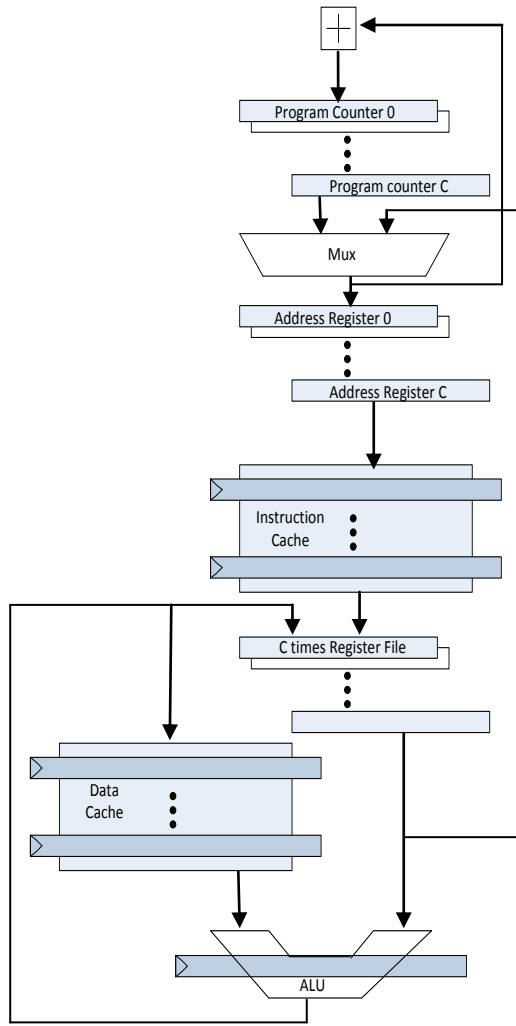

**Fig. 4. A generic C-slow processor Architecture**

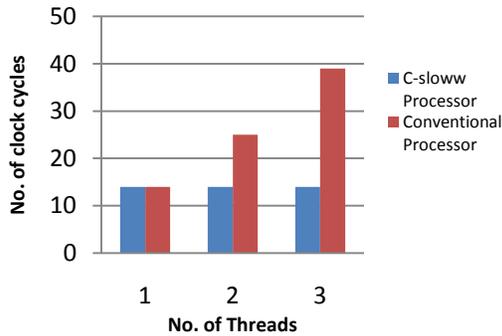

**Fig. 5 Comparison of number of clock cycles vs number of threads**

Table 1 represents *SPARTAN 3* FPGA device utilization summary of simple processor while fig.7 shows brief summary of same device utilization when *3-slow* design is implemented.

**Table 1 Device Utilization Summary of Simple Processor Design**

| Logic Utilization | Used | Available | Utilization |
|---|---|---|---|
| **No. of slice registers** | 2107 | 26,624 | 7% |
| **No. of 4 input LUTs** | 1444 | 26,624 | 5% |
| Logic Distribution | | | |
| **No. of occupied slices** | 1778 | 13,312 | 13% |
| **No. of slices containing only related logic** | 1778 | 1778 | 100% |
| **No. of slices containing only unrelated logic** | 0 | 1778 | 0% |
| Total No. of 4 input LUTs | 1454 | 26,624 | 5% |
| **No used as logic** | 1444 | | |
| **No used as a route-thru** | 10 | | |
| **No of bonded IOBs** | 9 | 333 | 2% |
| **No. of GCLKs** | 2 | 8 | 5% |

**Table 2 Device Utilization Summary of 3-Slow Processor Design**

| Logic Utilization | Used | Available | Utilization |
|---|---|---|---|
| **No. of slice registers** | 4,270 | 26,624 | 16% |
| **No. of 4 input LUTs** | 3,166 | 26,624 | 11% |
| Logic Distribution | | | |
| **No. of occupied slices** | 3,708 | 13,312 | 27% |
| **No. of slices containing only related logic** | 3,708 | 1778 | 100% |
| **No. of slices containing only unrelated logic** | 0 | 1778 | 0% |
| Total No. of 4 input LUTs | 3,167 | 26,624 | 11% |
| **No used as logic** | 3,167 | | |
| **No used as a route-thru** | 1 | | |
| **No of bonded IOBs** | 10 | 333 | 3% |
| **No. of GCLKs** | 2 | 8 | 25% |

After testing the simple microprogrammed FSM system and 3-slow on microprogrammed FSM system, comparison of different efficiency parameters between the said systems can be observed in table 3.

**Table 3 Different Efficiency Parameters Comparison between Simple and 3-slow Microprogrammed FSM**

| Efficiency Parameters | Microprogrammed FSM | 3-slow Microprogrammed FSM |
|---|---|---|
| Minimum period | 29.976ns | 11.558ns |
| Minimum input arrival time before clock | 29.985ns | 5.458ns |
| Maximum Frequency | 33.360MHz | 86.520MHz |
| Maximum output required time after clock | 7.165ns | 7.165ns |

Hence from the simulation results, it was observed that C-slowing is a promising technique to increase the performance of system especially when the system have to complete same task multiple times as in embedded systems.

## 5. CONCLUSION

From all previous discussion, from introduction to simulation results, one thing that is clear that for consumer electronics especially embedded system designs, SMT is the most promising way to enhance the performance. We have discussed many techniques to implement the SMT with different proposed models. One thing that is common in all these that they enhance performance by increasing number of processing units such as arithmetic and logic unit (ALU), floating point unit (FPU), and others that are the primary cause to increase the area of the system.

Most of embedded systems are area critical i.e. they require maximum performance with minimum area. To achieve this goal, we proposed C-slow processing technique. The most advantage of this technique, as discussed in simulation results, it use the available resources not from external resources and enhance the performance. From the simulation results, it is clear that 3-slow processor can compute 3 threads simultaneously i.e. producing the effect of three processing units are present in the design.

## 6. REFRENCES

[1] John Goodacre, Andrew N. Sloss, Parallelism and the ARM Instruction Set Architecture, IEEE Published by the IEEE Computer Society, 2005

[2] W. Knight, "Two Heads Are Better Than One", IEEE Review, September 2005

[3] D. Pham et al. The Design and Implementation of a First Generation CELL Processor. *In Proceeding of the IEEE International Solid-State Circuits Conference*, 2005.

[4] J. Cornish. Balanced Energy Optimization. In *International Symposium on Low Power Electronics and Design*, 2004

[5] A. Suga et al. FR-V Single-Chip Multicore Processor: FR1000. *Fujitsu Sci Tech J*, 42(2):190–199, 2006.

[6] Schlett M., "Trends in embedded-microprocessor design", *IEEE Computer*, pp. 44–49, Aug. 1998.

[7] "An Infrastructure for Designing Custom Embedded Counterflow Pipelines" Proceedings of the 33rd Hawaii International Conference on System Sciences – 2000

[8] D. M. Tullsen, S. J. Eggers, and H. M. Levy. Simultaneous Multithreading: Maximizing On-chip Parallelism. *Proceedings of the 22nd International Symposium on Computer Architecture*, pp. 206-218, June 1995.

[9] "Simultaneous Multithreading: a Platform for Next Generation Processors" Paulo Alexandre Vilarinho Assis *IEEE MICRO September/October 1997*

[10] Instruction Level Parallelism through Microthreading—A Scalable Approach to Chip Multiprocessors. The Computer Journal 2006 49(2). British Computer Society

[11] Instruction-Level Parallel Processing: History, Overview and Perspective *The Journal of Supercomputing,* Volume 7, No.1, January, 1993

[12] D.W. Wall, "Limits of Instruction-Level Parallelism," *Pruc. Fourth Int'l Con5 Architectural Support for Programming Languages and Operating Systems,* pp. 176-188, Apr. 1991.

[13] M. Johnson, "Super-scalar Processor Design," Technical Report No. CSL-TR-89-383, Stanford Univ., June 1989.

[14] K. Murakami, N. Irie, M. Kuga, and S. Tomita,"SIMP (Single Instruction Stream/Multiple Instruction Pipelining): A Novel High-speed Single-Processor Architecture," *Proc. 16th Ann. Int'l Symp. Computer Architecture,* pp. 78-85, May 1989.

[15] EXPLOITING INSTRUCTIONAND DATA-LEVEL PARALLELISM. *Roger Espasa Mateo Valero Polytechnic University of Catalunya-Barcelona* IEEE Micro *September/October 1997*

[16] *The Superthreaded Processor Architecture* Jenn-Yuan Tsai, Member, IEEE, Jian Huang, Student Member, IEEE, Christoffer Amlo, David J. Lilja, Senior Member, IEEE, and Pen-Chung Yew, Fellow, IEEE, IEEE TRANSACTIONS ON COMPUTERS, VOL. 48, NO. 9, SEPTEMBER 1999

[17] M.S. Schlansker and B.R. Rau, *EPIC: An Architecture for Instruction-Level Parallel Processors*, HPL Tech. Report HPL-1999-111, Hewlett-Packard Laboratories, Jan. 2000.

[18] IA-64 Application Developer's Architecture Guide, Intel Corp., 1999.

[19] M. S. Schlansker and B. R. R. Cover, "EPIC: Explicitly parallel instruction computing," *Computer*, vol. 33, no. 2, pp. 37–45, Feb. 2000.

[20] H. Sharangpani and K. Arora, "Itanium processor microarchitecture," *IEEE Micro*, vol. 20, pp. 24–43, Sept./Oct. 2000.

[21] SMTA: next-generation high-performance multi-threaded processor. J.-F. Tu and L.-H. Wang IEEE Proc.-Comput. Digit. Tech., Vol. 149, No. 5, September 2002

[22] K. Sankaralingam et al. , "Exploiting ILP TLP, and DLP with the Polymorphous TRIPS Architecture," *Proc. 30th Int'l Symp. Computer Architecture* (ISCA 03), ACM Press, 2003, pp. 422-433.

[23] DESIGN SPACE EXPLORATION FOR REAL-TIME EMBEDDED STREAM PROCESSORS Joseph R. Cavallaro Scott Rixner Rice University Sridhar Rajagopal WiQuest Communications IEEE MICRO JULY–AUGUST 2004

[24] Area and System Clock Effects on SMT/CMP Throughput James Burns and Jean-Luc Gaudiot, Fellow, IEEE, IEEE TRANSACTIONS ON COMPUTERS, VOL. 54, NO. 2, FEBRUARY 2005

[25] R. Bahr, S. Ciavaglia, B. Flahive, M. Kline, P. Mageau and D. Nickel. The DNl0000TX: a new high-performance PRISM processor. Proc. COMPCON '91 (1991). 7G. Blanck and S. Krueger. The SuperSPARCrM microprocessor. Proc. COMPCON '92 (1992), 136-141

[26] 2M. Johnson. Superscalar Microprocessor Design. (Prentice-Hall, Englewood Cliffs, New Jersey, 1991).

[27] N. P. Jouppi. The nonuniform distribution of instruction-level and machine parallelism and its effect on performance. IEEE Transactions on Computers C-38, 12 (December 1989), 1645-1658

[28] C. Leiserson, F. Rose, and J. Saxe, "Optimizing synchronous circuitry by retiming," *Proceedings of the 3rd Caltech Conference On VLSI*, pp. 87-116, March 1983.

[29] N. Weaver, Y. Markovskiy, Y. Patel and J. Wawrzynek, "Post placement c-slow retiming for the Xilinx Virtex FPGA," *Proceedings of the 11th ACM*

Symposium of Field Programmable Gate Arrays, Feb. 2003, pp. 185-194.

[30] R. Alverson, D. Callahan, D. Cummings, B. Koblenz, A. Porterfield, B. Smith. The Tera computer system. *Proceedings of the 1990 International Conference on Supercomputing*, 1990.

## 7. APPENDIX

**TABLE 4**
**Symbolic Instruction Set Architecture**

| Address | | Symbolic Instruction |
|---|---|---|
| 0 | | pc ← 0 |
| 1 | Fetch | MAR← pc |
| 2 | | IR ← M(MAR) ; pc ← pc+1 |
| 3 | Decode | I3 = 1? go to MEMREF |
| 4 | | XC0 = 1? Go to CMA |
| 5 | | XC1 = 1? Go to INCA |
| 6 | | XC2 = 1? Go to DCA |
| 7 | | go to HALT |
| 8 | CMA | A ← Ā |
| 9 | | go to Fetch |
| 10 | INCA | A ← A+1 |
| 11 | | go to Fetch |
| 12 | DCRA | A ← A-1 |
| 13 | | go to Fetch |
| 14 | MEMREF | if XC0 = 1, LDSTO |
| 15 | | if XC1 = 1, ADDSUB |
| 16 | | if XC2 = 1, JUMP |
| 17 | AND | MAR ← pc |
| 18 | | Buffer←M(MAR), pc← pc+1 |
| 19 | | MAR ← Buffer |
| 20 | | Buffer ← M(MAR) |
| 21 | | A ← A & Buffer |
| 22 | | go to Fetch |
| 23 | LDSTO | MAR ← pc |
| 24 | | Buffer ← M(MAR); pc ← pc +1 |
| 25 | | MAR ← Buffer |
| 26 | | if I0 = 1 go to STO |
| 27 | LOAD | Buffer ← M(MAR) |
| 28 | | A ← Buffer |
| 29 | | go to Fetch |
| 30 | STO | M(MAR) ← A |
| 31 | | go to Fetch |
| 32 | ADSUB | MAR ← pc |
| 33 | | Buffer←(MAR);pc← pc+1 |
| 34 | | MAR ← Buffer |
| 35 | | Buffer ← M(MAR) |
| 36 | | if I0 = , go to SUB |
| 37 | ADD | A ← A + Buffer |
| 38 | | go to Fetch |
| 39 | SUB | A ← A- Buffer |
| 40 | | go to Fetch |
| 41 | JUMP | MAR ← pc |
| 42 | | if I0 =0, go to JOZ |
| 43 | | if I0 =1, go to JOC |
| 44 | JOZ | if z=1 go to LOADPC |
| 45 | | pc ← pc+1 |
| 46 | | go to Fetch |
| 47 | JOC | if c=1 go to LOADPC |
| 48 | | pc ← pc+1 |
| 49 | | go to Fetch |
| 50 | LOADPC | pc ← M(MAR) |
| 51 | | go to Fetch |
| 52 | HALT | go to HALT |